\documentclass[aps,prl,reprint,twocolumn,showpacs]{revtex4}
\usepackage{epsfig}
\usepackage{subfigure}
\usepackage{graphicx}
\usepackage{dcolumn}
\begin{document}
\title{Generalized $X$ states of $N$ qubits and their symmetries}
\author{Sai Vinjanampathy}
\email[]{saiv@phys.lsu.edu}
\author{A. R. P. Rau}
\affiliation{Department of Physics and Astronomy, Louisiana State University, Baton Rouge, Louisiana, 70803-4001}
\date{\today}
\begin{abstract}
Several families of states such as Werner states, Bell-diagonal states and Dicke states are useful to understand multipartite entanglement. Here we present a $(2^{N+1}-1)$-parameter family of $N$-qubit ``$X$ states" that embrace all those families, generalizing previously defined states for two qubits. We also present the algebra of the operators that characterize the states and an iterative construction for this algebra, a sub-algebra of $su(2^{N})$. We show how a variety of entanglement witnesses can detect entanglement in such states. Connections are also made to structures in projective geometry.
\end{abstract}
\maketitle
Entanglement is a key feature of multipartite quantum systems and has been studied as a resource in varied fields such as computing \cite{nielsen2000quantum}, teleportation \cite{bennett1993teleporting}, metrology \cite{caves1981quantum}, secret sharing \cite{hillery1999quantum} and imaging \cite{PhysRevLett.74.3600}. Consequently, the characterization and the evolution of multipartite entanglement have generated a lot of interest in recent years.

One question concerns the transformation between any two multipartite states. This question is related to the number of entanglement classes that exist for a given $N$-qubit state. There is only one class of two-qubit states since it is well known that a Bell state can be probabilistically transformed to any two-qubit state via stochastic local operations and classical communication through which we define equivalence classes.   This is not true for a more general multipartite state. For instance, three-qubit states can be classified into four separate classes \cite{PhysRevLett.87.040401} and four-qubit states into nine \cite{4qubitClass}.

 For two qubits, the Werner state \cite{werner1989quantum} is an example of a one-parameter family that encompasses both separable and entangled states. The Bell-diagonal states are a three-parameter family of states which have maximally mixed marginals \cite{luo:042303}. All these are subsets of a seven-parameter family, called $X$ states, that occur in a variety of contexts such as entanglement and its decay under decoherence \cite{Yu_Eberly} and in describing other quantum correlations besides entanglement such as discord \cite{PhysRevA.81.042105}. They were defined \cite{Yu_Eberly} for two-qubit systems as states whose density matrix has non-zero elements only along its diagonal (three real parameters) and anti-diagonal (two complex parameters) in resemblance to the letter X. Recently \cite{rau2009algebraic}, an algebraic characterization was provided based on the symmetries of the sub-algebra of the states and operators involved. Extending this algebraic characterization to $N$-qubit $X$ states, we present several aspects of the algebra of the operators involved and some applications.

These alternative subsets of seven $X$ states describe a wide variety of physics in quantum information while still restricted to about half the number of parameters (7 \textit{vs.} 15) of the general two-qubit system. This restriction helps to calculate entanglement and other correlations analytically \cite{PhysRevA.81.042105,rau2008hastening}, allowing for more insight than numerical computations. It can be expected, therefore, that for $N$ qubits, with an exponential increase in the number of parameters, that similar subsets of $N$-qubit $X$ states with fewer parameters but still embracing most of the phenomena of interest will be worth studying. We develop such a description here.

 Before turning to geometric and group theoretic structures of the states and operators involved, note from the elementary viewing of the density matrix in the form of the letter X, that $2^{N}-1$ real parameters along the diagonal and $2^{N-1}$ complex values on the anti-diagonal add to a total of $2^{N+1}-1$ real parameters in the $X$ state. In terms of a $2^{N}$-level system in atoms, molecules or quantum optics, degeneracies and selection rules that restrict the couplings also lead to a consideration of such $X$ states.

A two-qubit system is the simplest model to study entanglement. Its 16 operators form a group under multiplication and commutation. A suitable representation involves Pauli matrices and was presented in \cite{PhysRevA.61.032301}. We will use the notation whereby $\sigma_{x}\otimes\tau_{z}$ is written as $X_{1}Z_{2}$. With this notation, two-qubit $X$ states are given by
\begin{eqnarray}\label{2_qubit_density}
\rho=\frac{1}{2^2}\sum_{i=0}^{2^2-1}(d_{i}\hat{D}_{i}+a_{i}\hat{A}_{i}).
\end{eqnarray}
Here $\hat{D}_{i}$ stands for the operator obtained by replacing 0 with $I$ and 1 with $Z$ in the binary rendering of $i$. The operator $\hat{A}_{i}$ is obtained by replacing similarly 0 with $X$ and 1 with $Y$. For example, since the number 2 is represented in binary as 10, we have $\hat{D}_{2}=I_{1}Z_{2}=Z_{2}$. Similarly, $\hat{A}_{2}=X_{1}Y_{2}$. Note that since Tr$(\rho)=1$, the coefficient $d_{0}=1$.

The two-qubit density matrix $\rho$ is Hermitian which implies that $\{a_i,d_i\}$ are real. Various choices of the coefficients lead to different states that are of interest. For instance, the Bell density matrix $\vert\Phi^{+}\rangle\langle\Phi^{+}\vert$ corresponds to  $d_{3}=a_{0}=1$, $a_{3}=-1$ and $d_{1}=d_{2}=a_{1}=a_{2}=0$. The Werner state \cite{werner1989quantum} is given by $\vert\psi\rangle=\frac{1-p}{4}I_1I_2+p\vert\Phi^+\rangle\langle\Phi^+\vert$. The choice $a_1=a_2=d_1=d_2=0$ corresponds to the general Bell-diagonal state, characterized by the three non-zero coefficients $(a_{0},a_{3},d_{3})$, and has been studied in the context of quantum correlations and decoherence \cite{luo:042303}.

Two-spin $X$ states arise in various physical systems. In \cite{PhysRevLett.87.050401}, the authors considered entanglement of an atom interacting with a quantized electromagnetic field. In \cite{sarandy:022108}, the author studied $X$ states in condensed matter systems for the role of quantum correlations in driving a quantum phase transition. In \cite{Yu_Eberly}, the authors studied the evolution of entanglement in $X$ states that were subject to spontaneous emission. They showed that $X$ states preserve their form under general forms of decoherence \cite{Yu_Eberly} and that some  disentangle at finite time. In \cite{rau2008hastening}, it was shown that this ``sudden death of entanglement" can be hastened, delayed or averted by using local operations.

  The seven operators involved in the definition of $\rho$ in Eq.(\ref{2_qubit_density}) have recently been identified \cite{rau2009algebraic} as  belonging to the $\mathsf{G}_{2}=$ SU(2)$\times$U(1)$\times$SU(2) subgroup of SU(4). One of the operators, namely $\hat{D}_{3}=Z_{1}Z_{2}$, is the U(1) element that commutes with the other six matrices and characterizes this set of $X$ states. Note also that the commuting operator is known as the ``stabilizer" of Bell states in quantum error correction \cite{stabilizer}. In a series of papers \cite{rau2009algebraic}, it was established that the multiplication table of the seven operators involved can be represented by a triangle (Fig. 1 of the first reference in \cite{rau2009algebraic}). The seven operators are associated with the three vertices, three midpoints of sides and the incenter of the triangle. In projective geometry, this diagram describes the smallest projective plane called the Fano plane PG(2,2), with a complete duality of seven points incident on seven lines (including the in-circle) \cite{coxeter1973regular}.
  
   We will use the notation from convex geometry whereby the $m$-face of an $N$-dimensional polytope refers to an $m$-dimensional sub-polytope \cite{coxeter1973regular}. Also, the $N$-simplex is an $N$-dimensional polytope with $N$+1 vertices, an example of which in two dimensions is the triangle. With this notation, we see that the operators of $\mathsf{G}_2$ are associated with three 0-faces (vertices), three 1-faces (edges) and one 2-face (the ``face" of the triangle) of a 2-simplex. These definitions will be generalized to understand the diagram related to $N$-qubit $X$ states in the next section.
Motivated by the two-qubit $X$ state, we introduce the $N$-qubit generalization of $X$ states as
\begin{eqnarray}
\rho=\frac{1}{2^N}\sum_{i=0}^{2^N-1}(d_{i}\hat{D}_{i}+a_{i}\hat{A}_{i}).
\end{eqnarray}
As before, $\{a_{i},d_{i}\}$ are real and $d_0=1$. Note that the commuting elements for the $N$-qubit $X$ state are given by $N\choose2$ operators $Z_{i}Z_{j}$ where $i\neq j$ and $i,j=1 \ldots N$, plus $N\choose4$ quadruple products $Z_{i}Z_{j}Z_{k}Z_{l}$, etc., for a total of $2^{N-1}-1$ U(1) operators. The larger set of $2^N$ operators $\hat{D}_{i}$ that includes all products of $Z_{i}$ commute with each other, but not with all the operators $\hat{A}_{i}$ as do the U(1) elements. The invariance group for $N$-qubit $X$ states $\mathsf{G}_{N}$ is iteratively constructed from that of the ($N$-1)-qubit $X$ state by concatenation: $\mathsf{G}_{N}=\mathsf{G}_{N-1}\times$U(1)$\times\mathsf{G}_{N-1}$. For example, $\mathsf{G}_{1}$ is the SU(2) group of a single qubit $X$ state , the two-qubit $X$ state is given by $\mathsf{G}_{2}=$SU(2)$\times$U(1)$\times$SU(2), and the three-qubit $X$ state is given by $\mathsf{G}_{3}$=SU(2)$\times$U(1)$\times$SU(2)$\times$U(1)$\times$SU(2)$\times$U(1)$\times$SU(2) consisting of 15 operators, three of them, the U(1) elements $Z_{i}Z_{j}$, commuting with every member of the set of fifteen. $\mathsf{G}_{N}$ includes $2^{N-1}$ SU(2)s in its total of $2^{N+1}-1$ operators. In projective geometry, it corresponds to PG($N$,2), generalizing the Fano plane for two qubits. In the related subject called design theory \cite{beth1999design}, it is called a $2-(2^{N+1}-1,3,1)$ design.

\begin{figure}
     \centering
     \subfigure[]{
          \label{tetra}
          \includegraphics[width=.42\textwidth]{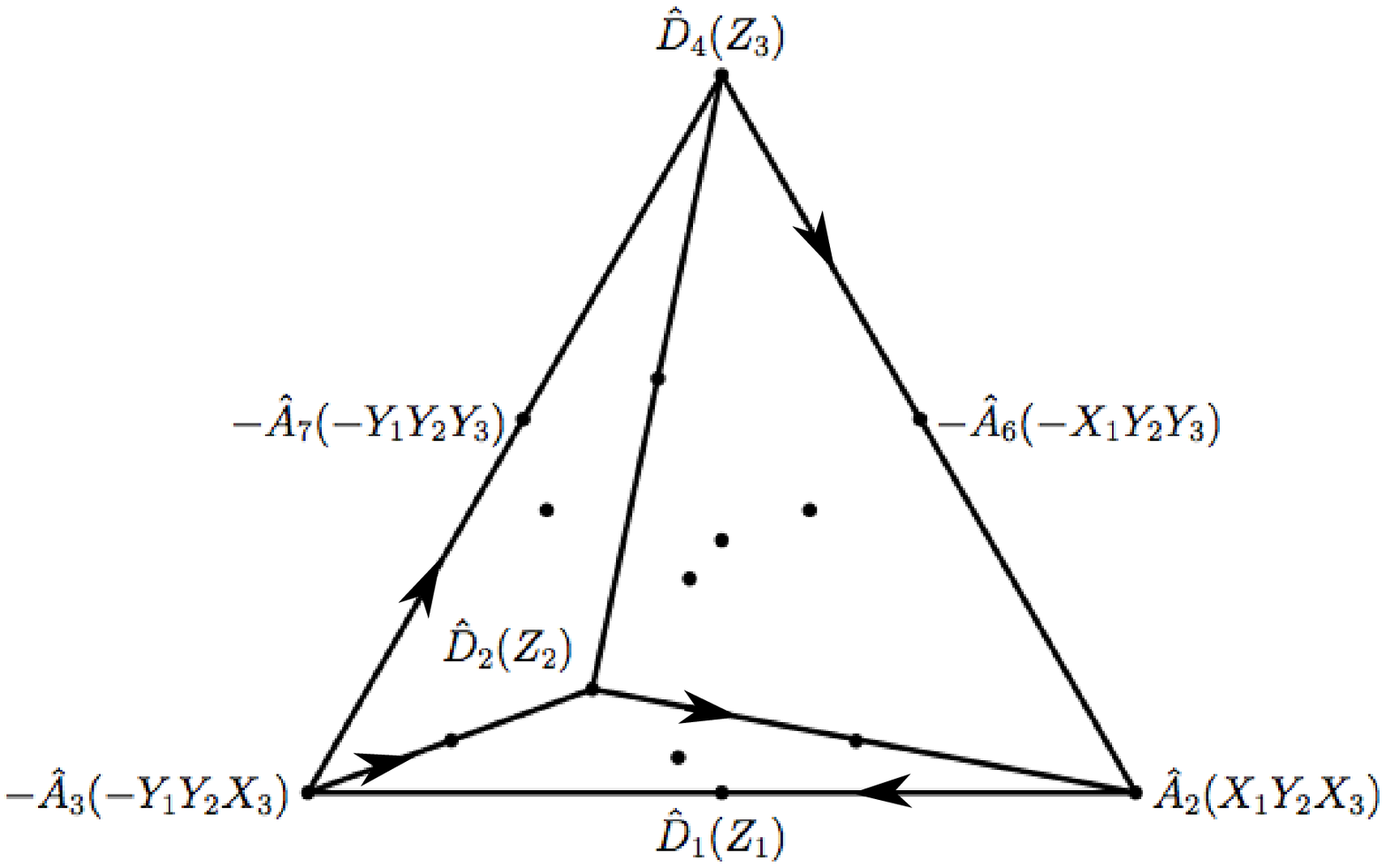}}
     \vspace{.2in}
     \subfigure[]{
           \label{planar}
           \includegraphics[width=.39\textwidth]
                {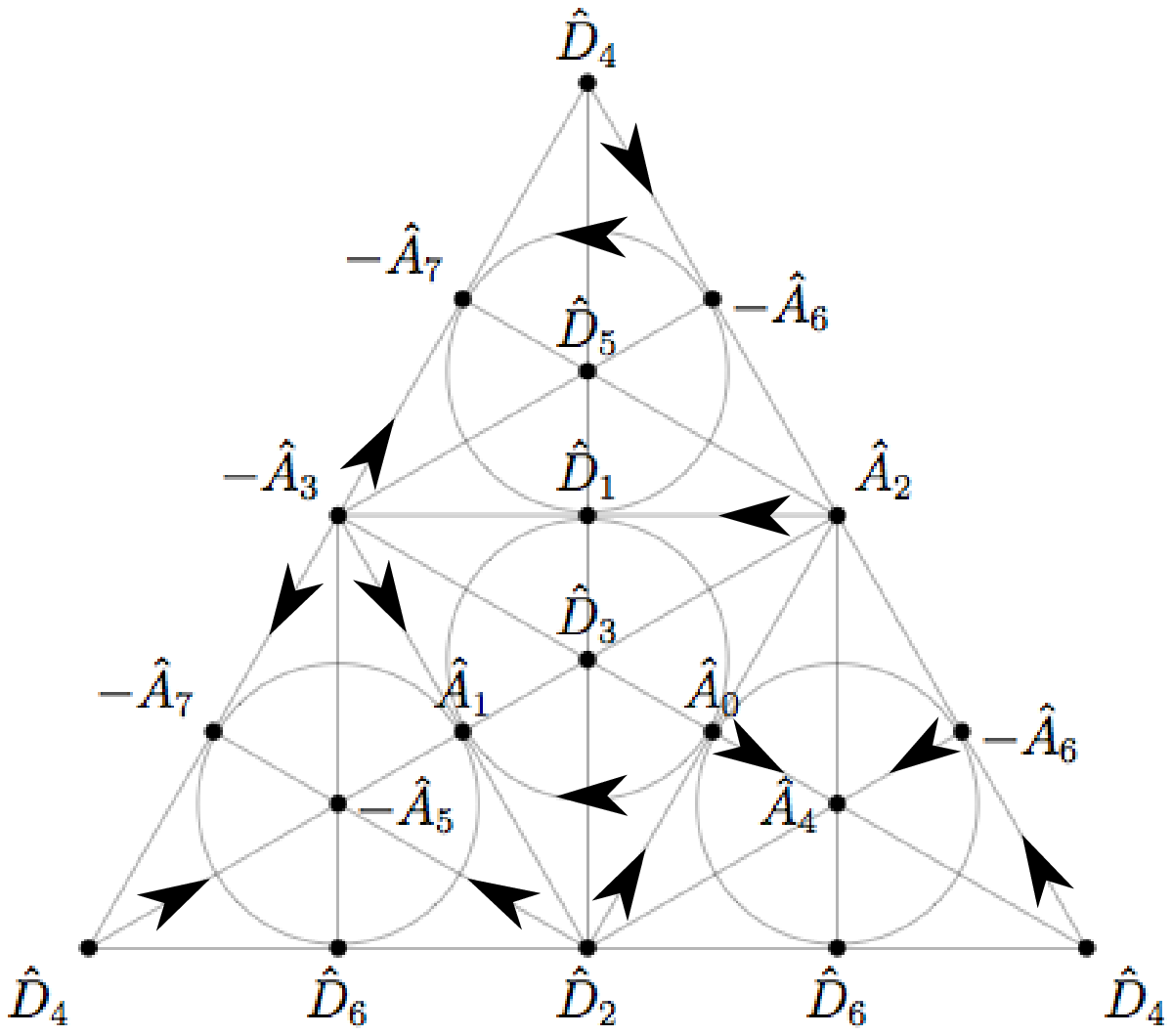}}
     \caption{The tetrahedron or 3-simplex associated with three-qubit $X$ states. In (a), the 15 operators in Eq. (\ref{3_qubit_density}) are identified with the points of the simplex(vertices, mid-points of edges, face centers and body center of the tetrahedron). For clarity, only a few points are labeled. The same tetrahedron is opened out into a planar diagram in (b), resulting in the vertex associated with $\hat{D}_4$ repeated three times. Six lines connecting pairs of face centers and all seven lines through the body center are omitted for clarity. Arrowed lines connecting three operators denote that the product of any two gives the third operator in a cyclic fashion, with a multiplicative $\pm i$. Unarrowed lines denote the product of any two as the third operator, regardless of the order.}
\label{3q}
\end{figure}

 This approach extends to the geometry of the operators involved. First, a general single qubit state is trivially an $X$ state. The three Pauli operators involved in defining this state (besides the unit operator) can be associated with the two endpoints $X$ and $Y$ of a line and the center $Z$ of the line: see bottom line of Fig. \ref{tetra}. Such a line is a 1-simplex, whose two 0-faces and one 1-face are associated with the operators involved in defining a single-qubit $X$ state. Next, the triangle involved in defining the two-qubit $X$ state can be thought of as the addition of a 0-face (third vertex). The addition of this 0-face $Z_{2}$, and simultaneously multiplying by $Y_2$ (or alternatively $X_2$) the end-points of the initial 1-simplex that forms the base of the triangle, brings in two additional 1-faces (vertices) and one 2-face (in-center $Z_{2}Z_{1}$). The density matrix of the $X$ state can now be written as a sum over the seven $4\times4$ matrices as noted in \cite{rau2009algebraic}. Fig. 1 of that reference, now incorporated as the base of the tetrahedron in Fig. \ref{tetra}, renders compactly the states, operators and multiplications between them so that all manipulations and calculations of two-qubit $X$ states reduce to inspection.
 
 In this manner, the operators $\{\hat{D}_{i},\hat{A}_{i}\}$ of an $N$-qubit $X$ state can be constructed by adding a 0-face to the ($N$-1)-simplex describing the ($N$-1)-qubit $X$ state. The number of $m$-faces of an $N$-simplex is given by ${N+1\choose m+1}$. The sum of all $m$-faces for $m \leq N$ is $\sum_{m=0}^{N+1}=2^{N+1}-1$. These number counts agree with the ones given above of the SU(2) and U(1) operators. Hence, we can associate the states and operators of generalized $X$ states with the $m$-faces of an $N$-simplex.

As an example, we consider in detail the three-qubit $X$ states, written explicitly as
\begin{widetext}
\begin{eqnarray}\label{3_qubit_density}
\rho=
\frac{I_{1}I_{2}I_{3}}{8}+\frac{1}{8}(d_{1} Z_{1}+d_{2} Z_{2}+d_{3} Z_{1}Z_{2}+d_{4}Z_{3}+d_{5}Z_{1}Z_{3}+d_{6}Z_{2}Z_{3}+d_{7}Z_{1}Z_{2}Z_{3})\nonumber\\
+\frac{1}{8}(a_{0}X_{1}X_{2}X_{3}+a_{1}Y_{1}X_{2}X_{3}+a_{2}X_{1}Y_{2}X_{3}
+a_{3}Y_{1}Y_{2}X_{3}+a_{4}X_{1}X_{2}Y_{3}+a_{5}Y_{1}X_{2}Y_{3}+a_{6}X_{1}Y_{2}Y_{3}+a_{7}Y_{1}Y_{2}Y_{3}).
\end{eqnarray}
\end{widetext}
The 15 operators involved may be identified with the four vertices, mid-points of six edges, four face-centers and one body center of a tetrahedron. This diagram is given in Fig. \ref{3q}.

As in the case of two-qubit $X$ states, different choices of the parameters $\{a_i,b_i\}$ lead to different states that are of physical interest. The choice of $d_{4}=1=d_{5}=d_{6}=a_{0}=1$, $a_{3}=a_{5}=a_{6}=-1$ and the other parameters equal to zero corresponds to a GHZ state \cite{GHZ}. Tracing over any subsystem of this density matrix yields a completely mixed state. 

There are three commuting elements now, namely $Z_{1}Z_{2}$, $Z_{2}Z_{3}$ and $Z_{1}Z_{3}$,  instead of just one for a two-qubit $X$ state. Any two of these are the so-called stabilizers of the GHZ state (since the product of any two is the third operator). As with two qubits \cite {rau2009algebraic}, other choices of commuting operators yield different classes of tripartite $X$ states. Specifically, the choice of $Y_{1}Y_{2}$, $Y_{2}Y_{3}$ and $Y_{1}Y_{3}$ yields an $X$ state all of whose elements are non-zero, a generalization of a similar two-qubit example in \cite{rau2009algebraic}. This matrix is explicitly written as
\begin{widetext}
\begin{equation}\label{3Xbasis}
\rho=\frac{1}{8}
\begin{small}
\left(
\begin{array}{cccccccc}
 1+a_0 & a_1-i d_1 & a_2-i d_2 & a_3-d_3 & a_4-i d_4 & a_5-d_5 & a_6-d_6 & a_7+i d_7 \\
 a_1+i d_1 & 1-a_0 & a_3+d_3 & -a_2-i d_2 & a_5+d_5 & -a_4-i d_4 & a_7-i d_7 & -a_6-d_6 \\
 a_2+i d_2 & a_3+d_3 & 1-a_0 & -a_1-i d_1 & a_6+d_6 & a_7-i d_7 & -a_4-i d_4 & -a_5-d_5 \\
 a_3-d_3 & -a_2+i d_2 & -a_1+i d_1 & 1+a_0 & a_7+i d_7 & -a_6+d_6 & -a_5+d_5 & a_4-i d_4 \\
 a_4+i d_4 & a_5+d_5 & a_6+d_6 & a_7-i d_7 & 1-a_0 & -a_1-i d_1 & -a_2-i d_2 & -a_3-d_3 \\
 a_5-d_5 & -a_4+i d_4 & a_7+i d_7 & -a_6+d_6 & -a_1+i d_1 & 1+a_0 & -a_3+d_3 & a_2-i d_2 \\
 a_6-d_6 & a_7+i d_7 & -a_4+i d_4 & -a_5+d_5 & -a_2+i d_2 & -a_3+d_3 & 1+a_0 & a_1-i d_1 \\
 a_7-i d_7 & -a_6-d_6 & -a_5-d_5 & a_4+i d_4 & -a_3-d_3 & a_2+i d_2 & a_1+i d_1 & 1-a_0
\end{array}
\right).
\end{small}
\end{equation}
\end{widetext}
Tracing over any one of the qubits now yields a reduced density matrix whose coherences are non-zero unlike in the previous paragraph. We will return to the importance of this result below.

Consider a GHZ state shared between three parties, Alice, Bob and Charlie. A GHZ state $(\vert 000\rangle +\vert 111\rangle)/\sqrt{2}$ that is subject to a fairly general model of decoherence ( such as amplitude damping, phase damping, or spontaneous emission) involves all the operators in Eq. (\ref{3_qubit_density}) and hence evolves as a three-qubit $X$ state. Alternatively, a GHZ state may be defined as $(\vert +++\rangle +\vert ---\rangle)/\sqrt{2}$, where $\vert\pm\rangle=(\vert0\rangle\pm\vert 1\rangle)/\sqrt{2}$. While the first definition corresponds to the commuting elements $Z_iZ_j$, the latter definition corresponds to the commuting elements $X_iX_j$(a similar result pertains to $Y_{i}Y_{j}$). Note that the two definitions of the GHZ state are related by local unitary transformations. If the qubit held by Alice is now traced over, it can be verified that the remaining two-qubit state has no entanglement. But, as the coherences of the two-qubit density matrix are non-zero, there are non-classical correlations that are present between Bob and Charlie that can be quantified by a measure of quantum correlations such as quantum discord \cite{henderson2001classical}.  Hence $X$ states characterized by different commuting elements can have drastically different correlation properties in their marginals. These non-classical correlations may provide speedup for certain tasks \cite{PhysRevLett.81.5672}.

 To detect different types of entanglement in three and four qubits, we consider witness operators that detect GHZ-type entanglement, W-type entanglement and the witness corresponding to symmetric Dicke states \cite{Guhne20091} for $N=3,4$ qubits. $N$-qubit $X$ states characterized by products of $Z$ operators as in Eq.(\ref{3_qubit_density}) are readily seen to possess GHZ-type entanglement  for arbitrary N. Consider an $X$ state characterized by products of $X$ operators with $a_0=a_3=a_5=a_6=d_1=d_2=d_4=d_7=0$ and $-a_1=-a_2=-a_4=a_7=d_3=d_5=d_6=1$. For this state, $Tr(W_3\rho)=3/4$ where $W_3$ is the three-qubit W state. Thus the witness operator $2I/3-W_{3}$ detects W-type entanglement in this state. Furthermore, the four-qubit $X$ state characterized by products of $X$ operators with $d_1=d_2=d_4=d_7=d_8=d_{11}=d_{13}=d_{14}=a_1=a_2=a_4=a_7=a_8=a_{11}=a_{13}=a_{14}=0$ and $d_3=d_5=d_6=d_9=d_{10}=d_{12}=d_{15}=a_0=-a_3=-a_5=-a_6=-a_9=-a_{10}=-a_{12}=a_{15}=1$ is a state with $\langle D_{2,4}\vert \rho \vert D_{2,4}\rangle=3/4$. Here $\vert D_{2,4}\rangle$ is the symmetric Dicke state and $2I/3-\vert D_{2,4}\rangle\langle D_{2,4}\vert$ detects entanglement of the symmetric Dicke type in the given four-qubit $X$ state.

In summary, we have introduced a family of states called $X$ states for $N$ qubits analogous to those discussed for $N=2$, and have characterized them by a set of commuting operators. The algebra of the operators involved defines the family of states and also serves to describe operations on them. We have also presented a scheme for this algebra in terms of $N$-simplexes. Various entanglement witnesses were shown to detect entanglement in these states.

\textit{Note added.}--After completion of this work, a paper which deals with an example of  three-qubit X-states was posted \cite{weinstein}.


\begin{thebibliography}{45}
\bibitem{nielsen2000quantum}M. A. Nielsen and I. L. Chuang, \textit{Quantum computation and information} (Cambridge University Press, Cambridge, UK, 2000). 
\bibitem{bennett1993teleporting} C. Bennett, G. Brassard, C. Crepeau, R. Jozsa, A. Peres, and W. Wootters, Phys. Rev. Lett. \textbf{70}, 1895 (1993). 
\bibitem{caves1981quantum} C. Caves, Phys. Rev. D \textbf{23}, 1693 (1981). 
\bibitem{hillery1999quantum} M. Hillery, V. Bu\v{z}ek, and A. Berthiaume, Phys. Rev. A \textbf{59}, 1829 (1999). 
\bibitem{PhysRevLett.74.3600} D. V. Strekalov, A. V. Sergienko, D. N. Klyshko, and Y. H. Shih, Phys. Rev. Lett. \textbf{74}, 3600 (1995); M. I. Kolobov, Rev. Mod. Phys. \textbf{71}, 1539 (1999). 
\bibitem{PhysRevLett.87.040401}A. A\'{c}in, D. Bru\ss{}, M. Lewenstein, and A. Sanpera, Phys. Rev. Lett. \textbf{87}, 040401 (2001). 
\bibitem{4qubitClass} F. Verstraete, J. Dehaene, B. De Moor, and H. Verschelde, Phys. Rev. A \textbf{65}, 052112 (2002). 
\bibitem{werner1989quantum} R. Werner, Phys. Rev. A \textbf{40}, 4277 (1989). 
\bibitem{luo:042303} S. Luo, Phys. Rev. A \textbf{77}, 042303 (2008). 
\bibitem{Yu_Eberly} T. Yu and J. H. Eberly, Phys. Rev. Lett. \textbf{93}, 140404 (2004); Science \textbf{323}, 598 (2009).
\bibitem{PhysRevA.81.042105} M. Ali, A. R. P. Rau, and G. Alber, Phys. Rev. A \textbf{81}, 042105 (2010); J. Maziero, L. C. Celeri, R. M. Serra, and V. Vedral, \textit{ibid.} \textbf{80}, 044102 (2009). 
\bibitem{rau2009algebraic} A. R. P. Rau, J. Phys. A \textbf{42} 412002 (2009);Phys. Rev. A \textbf{79}, 042323 (2009); J. of Biosci. \textbf{34}, 353 (2009).
\bibitem{rau2008hastening} A. R. P. Rau, M. Ali, and G. Alber, Europhys. Lett. \textbf{82}, 40002 (2008); M. Ali, G. Alber, and A. R. P. Rau, J. Phys. A \textbf{42}, 025501 (2009).
\bibitem{PhysRevA.61.032301} A. R. P. Rau, Phys. Rev. A \textbf{61}, 032301 (2000).
\bibitem{PhysRevLett.87.050401} S. Bose, I. Fuentes-Guridi, P. L. Knight, and V. Vedral, Phys. Rev. Lett. \textbf{87}, 050401 (2001). 
\bibitem{sarandy:022108} M. S. Sarandy, Phys. Rev. A \textbf{80}, 022108 (2009). 
\bibitem{stabilizer} D. Gottesman, Phys. Rev. A \textbf{54}, 1862 (1996). 
\bibitem{coxeter1973regular} H. Coxeter, Duke Math. J \textbf{13}, 561 (1946).
\bibitem{beth1999design} T. Beth, D. Jungnickel, and H. Lenz, \textit{Design theory} (Cambridge University Press, Cambridge, UK, 1999). 
\bibitem{GHZ}D. M. Greenberger, M. A. Horne, A. Shimony, and A. Zeilinger, Am. J. Phys. \textbf{58}, 1131 (1990). 
\bibitem{henderson2001classical} L. Henderson and V. Vedral, J. Phys. A \textbf{34}, 6899 (2001); H. Ollivier and W. H. Zurek, Phys. Rev. Lett. \textbf{88}, 017901 (2001). 
\bibitem{PhysRevLett.81.5672} E. Knill and R. Laflamme, Phys. Rev. Lett. \textbf{81}, 5672 (1998); B. P. Lanyon, M. Barbieri, M. P. Almeida, and A. G. White, \textit{ibid.} \textbf{101}, 200501 (2008); G. Passante, O. Moussa, C. A. Ryan, and R. Laflamme, \textit{ibid.} \textbf{103}, 250501 (2009). 
\bibitem{Guhne20091} O. Guhne and G. Toth, Phys. Rep. \textbf{474}, 1 (2009).
\bibitem{weinstein}Y. S. Weinstein,e-print arXiv:1004.3748v1 [quant-ph].
\end{thebibliography}
\end{document}